\documentclass[10pt]{revtex4}
\usepackage{graphicx}
\usepackage{amsthm} 

\begin{document}

\title{Using Temporal Entanglement to Perform Thermodynamical Work}
\author{Vlatko Vedral}
\affiliation{Clarendon Laboratory, University of Oxford, Parks Road, Oxford OX1 3PU, United Kingdom\\Centre for Quantum Technologies, National University of Singapore, 3 Science Drive 2, Singapore 117543\\
Department of Physics, National University of Singapore, 2 Science Drive 3, Singapore 117542}

\begin{abstract}
Here we investigate the impact of temporal entanglement on a system's ability to perform thermodynamical work. We show that while the quantum version of the Jarzynski equality remains satisfied even in the presence of temporal entanglement, the individual thermodynamical work moments in the expansion of the free energy are, in fact, sensitive to the genuine quantum correlations. Therefore, while individual moments of the amount of thermodynamical work can be larger (or smaller) quantumly than classically, when they are all combined together into the (exponential of) free energy, the total effect vanishes to leave the Jarzynski equality intact. Whether this is a fortuitous coincidence remains to be seen, but it certainly goes towards explaining why the laws of thermodynamics happen to be so robust as to be independent of the underlying micro-physics. We discuss the relationship between this result and thermodynamical witnesses of spatial entanglement as well as explore the subtle connection with the ``quantum arrow of time".  
\end{abstract}

\maketitle

\section{Introduction}

Thermodynamics is such a fundamental theory that it is frequently said to be the only ``classical" theory to have withstood the onslaught of quantum physics. While Newtonian physics and classical electrodynamics both had to be modified to accommodate the basic notions of quantum physics, the first and the second law of thermodynamics remained completely unchanged. This said robustness of thermodynamics is, in some sense, both true and false. It is indeed correct that the formulation of the laws of thermodynamics we currently use is the same as that offered by Planck in a book completed before his quantum hypothesis paper. Moreover, Planck was even guided by classical thermodynamics to derive the black body formula that started the whole quantum revolution (this is precisely why we think of thermodynamics as more fundamental than any microscopic physics). On the other hand, it is not entirely true that thermodynamics remained unchanged with the advent of quantum physics. Thermodynamical notions of heat, work, energy and entropy do have to be adapted to quantum physics and some might say that this is still an unfinished endeavour as there is no consensus on exactly how quantisation of thermodynamics ought to be accomplished \cite{Hanggi,Koji-RMP}. 

In quantum mechanics states of systems are described using density matrices (instead of probabilities), which can lead to some seemingly paradoxical thermodynamical notion such as that of negative (conditional) entropy \cite{del Rio}. Furthermore, energy happens to be an operator in quantum physics (the Hamiltonian), while work (classically also a form of energy) does not seem to allow an operator description when translated into quantum physics \cite{Hanggi}. These differences might contribute to explaining why there are so many different approaches to quantum thermodynamics \cite{Mahler}. 

Making connections between quantum physics and thermodynamics, no matter how much this might be fraught with difficulties, is an important enterprise. This is for two reasons. One is that thermodynamical engines have entered the nano-domain where the laws of quantum physics can no longer be ignored. Therefore, in order to even be able to understand, formulate and exploit nano-machines we need to be able to phrase quantum thermodynamics consistently. The second point is less practical, but nonetheless equally important in the eyes of the present author. Thermodynamical laws appear to us not only well grounded in experiment, but also intuitive and natural. Quantum physics, despite all its experimental success, frequently doesn't. If we could show that all quantum quirks (such as the notions of superposition, non-locality and entanglement) can in a way be used to do something physically useful (such as thermodynamical work) then this might help us better understand, accept, and maybe even demystify quantum physics. 

The current work is motivated by exactly this demystifying quality of doing quantum thermodynamics. In order to make this paper accessible to a wide audience, we will first describe the notion of temporal quantum entanglement. This can be though of in many different ways, but the crux in all of them is that correlations between measurements at two different times, performed on the same quantum system (in this paper, almost always, a qubit), can be as high as the correlations that lead to the violation of Bell's inequalities for two qubits. The latter is known as spatial entanglement and hence the former is appropriately named temporal entanglement \cite{Vedral}. It has already been shown \cite{Maruyama} that spatial entanglement, just like other forms of correlation, can be used to extract work. Here we would like to show that temporal entanglement is also useful in the same thermodynamical sense. The main surprise will be that the Jarzynski equality \cite{Jarzynski}, relating average work to free energy (and also proven here for convenience), remains true notwithstanding temporal entanglement.
       
The paper is structured as follows. We will first derive the temporal Bell inequality \cite{Leggett,Vedral}, which was recently experimentally tested in \cite{Pryde}. Then we will present the proof, due to Tasaki \cite{Tasaki}, of the quantum version of the Jarzynski equality \cite{Jarzynski}. Finally, we show how to put the two results together in order to explain why quantum physics allows us, under some well defined conditions, to do more work than classically possible (cf \cite{Hide}). This conclusion will illustrate the point we made at the beginning: thermodynamics is and isn't the same after being quantized. Yes, the Jarzynski equality is valid in both classical and quantum thermodynamics and, in particular, is impervious to quantum entanglement (both spatial and temporal as will be seen). But, no, it does not lead to the same amount of work as far as its moments are concerned: quantum thermodynamics allows us to do more work than classically possible.

\section{Derivation of temporal Bell's inequalities}

Imagine an ensemble of quantum two level systems independently prepared in an identical fashion to be in some quantum state
$\rho$. Imagine furthermore that at first we have a choice between performing two dichotomic measurements (with outcomes plus and minus $1$), 
$A_1$ and $A_2$ on this ensemble which is then followed by either of the two dichotomic measurements $B_1$ and $B_2$. 
This is conducted in the following fashion. The first qubit is measured first and a fair coin is tossed to decide between $A_1$ and $A_2$. Then 
a coin is tossed again to decide between $B_1$ and $B_2$ and the resulting measurement is also executed on the first qubit. Once this is performed, we
move onto the second qubit, and repeat the same procedure. We continue until the whole ensemble is exhausted (on average therefore, we will have a quarter of the ensemble measuring each of the correlations $A_1B_1$, $A_1B_2$, $A_2B_1$ and $A_2B_2$). The procedure is the same as for ordinary spatial Bell's inequalities, the only difference being that the two measurements (the earlier and later) are both performed on the same quantum system (and not on two spatially separated ones). 

The point of this experiment will be to compute the following average value (which is the same as in the case of the Clauser-Horne-Shimony-Holt version of Bell's inequalities):
\begin{equation}
\langle {\cal B} \rangle := \langle A_1B_1 + A_2B_1 + A_1B_2 - A_2B_2\rangle \; .
\end{equation}
Quantum mechanically, this combination of measurements will have a maximum value given by $2\sqrt{2}$, again the same as in the spatial CHSH version of Bell's inequalities. The proof of this (perhaps at first sight surprising) fact is simple and it is instructive to go through it to see exactly how quantum physics achieves this particular value. Take the following choice of operators, $A_1 = Z$, $A_2 = (Z+X)/\sqrt{2}$, $B_1 = Z$ and $B_2 = (Z-X)/\sqrt{2}$ (these again are the operators that would lead to the maximal violation of the spatial CHSH inequality).
We then have  
\begin{equation}
\langle {\cal B} \rangle  = \sqrt{2} (XX +ZZ) \; .
\end{equation}
The average value of $XX$ is $1$, since a measurement of $X$ always leaves the system in an eigenstate of $X$, so that the
subsequent measurement of $X$ is always perfectly correlated with it. Likewise for the $ZZ$ measurement. Interestingly, this is
completely independent of the initial state $\rho$ (as noted in \cite{Vedral}) since the correlation between initial and final measurement
is perfect for any outcome and therefore the probabilities for different outcomes are immaterial. The independence of the initial state
holds for any choice of measurements and to see why we express each of the observables in terms of their Bloch representation, $A_{1,2} = a_{1,2}\cdot \sigma$
and $B_{1,2} = b_{1,2}\cdot \sigma$, where $c\cdot\sigma := c_x \sigma_x + c_y \sigma_y + c_z \sigma_z$. Then, $\langle AB\rangle = a\cdot b$ (where again $a\cdot b := a_x b_x + a_y b_y + a_z b_z$), which is already independent of the input state, so that the whole Bell average becomes
\begin{equation}
\langle A_1B_1 + A_2B_1 + A_1B_2 - A_2B_2\rangle = a_1\cdot b_1 + a_1\cdot b_2 + a_2\cdot b_1 - a_2\cdot b_2\; .
\label{quantum-Bell}
\end{equation}
We have therefore arrived at the conclusion that the maximum quantum value of temporal Bell inequalities is 
\begin{equation}
\langle \sqrt{2} (XX +ZZ)\rangle = 2\sqrt{2} \; .
\end{equation}
Classically, on the other hand, (here by classically we mean that $A$s and $B$s are just numbers, whose values are either $1$ or $-1$ and 
are independent of other measurements performed) the absolute value of $\langle A_1B_1 + A_2B_1 + A_1B_2 - A_2B_2\rangle$ can never exceed $2$. 
A way of understanding the difference between the quantum and classical averages is to view it within the consistent histories approach. Namely, classically, a
system can have one of a number of mutual exclusive (i.e. orthogonal) histories, while quantumly a system can exist in a ``superposition of histories". In other
words, in the classic version of the double slit experiment a particle has either gone through one or the other slit (two possible histories), but quantumly we know that the particle can also go through both slits at the same time. The philosophical implications of this have been extensively discussed elsewhere \cite{Leggett,Vedral} and our purpose here is different. We intend to use the quantum-classical gap to perform thermodynamical work. 
Before we do so, let us note the curious temporal symmetry in the above analysis. It is clear that the two times at which measurements are made on the qubit can easily be interchanged without changing the result (this is why our expression $\langle XX\rangle$ did not require any time label). Quantum physics is, in this sense, blind to the arrow of time despite the fact that performing a measurement is, following Bohr, frequently said to be an ``irreversible act of amplification". However, the symmetry of quantum measurements is not a surprise and it has been known for a long time \cite{Aharonov} that quantum measurements do not introduce any irreversibility into microscopic physics. With this in mind we turn to the quantum Jarzynski equality.   

\section{Derivation of the quantum Jarzynski relation}

The second law of thermodynamics states, in one of its many equivalent formulations, that the amount of work between two states cannot exceed the free energy difference between those states. The equality is reached only if we perform all operations in a reversible manner, in which case we can recover the maximum amount of work (equal to the free energy difference). However, the second law does not stipulate how much less we will get in any other particular irreversible setting. Interestingly enough, thermodynamical work and free energy can still be related through an equality, which is a result due to Jarzynski\cite{Jarzynski}. 

The Jarzynski relation states that 
\begin{equation}
\langle e^{\beta (W -\Delta F)} \rangle = 1
\end{equation}
where $W$ is the work the system does between the initial and final equilibrium states whose free energy difference is $\Delta F$. 
The brackets $\langle . \rangle$ indicate an average value of the quantity where the averaging is over all possible trajectories between the initial and the final state (and it will be defined below in the proof of the Jarzynski relation). The free energy is defined as $F=-kT\ln Z$, where $Z=\sum_n e^{-\beta E_n}$ is the partition function. 

\begin{proof}[{\bf Proof of Jarzynski equality}] 
Let us imagine the following protocol. We again have an ensemble of quantum systems (not necessarily qubits, though later we will specialize to them), each starting in a thermal state $\rho_i$, whose
corresponding partition function is $Z_i = tr \exp\{-\beta H_i\}$, where $H_i$ is the Hamiltonian. The following sequence of operations is now made on each of the systems. Firstly, a measurement is performed in the eigen-basis of $\rho_i$, labeled $P^i_n$ (which is the same as the eigen-basis of $H_i$). This is followed by a unitary transformation of the form $U = e^{-iH_f t}$. Finally a measurement is made in the basis of the Hamiltonian $H_f$, which we label $P^f_m$. It is immaterial what happens to the system beyond this point (in practice, it would typically thermalise into some state depending on its contact with the final environment). Note that the case of evolution analysed here is effectively the sudden quench scenario. The Jarzynski equality is much more general (it can include open system's evolution and feedback), but we need not consider this greater generality for our present purposes (for a general discussion see, for instance, \cite{Sagawa}).    

To prove Jarzynski's equality, we first define probabilities for each sequence of outcomes. More specifically, the probability that the first measurement yields the energy value $E^i_n$, while the second one yields energy eigenstate $E^f_m$ is
\begin{equation}
p(n,m) =  tr  \{ P^f_m U P^i_n\rho_0 P^i_n U^{\dagger} P^f_m \} = tr  \{ P^f_m P^i_n\rho_i \} =  tr  \{ P^f_m P^i_n\} \frac{e^{-\beta E^i_n}}{Z_i} \; ,
\end{equation}
where to derive this result we have used the fact that $P^f_m$ commute with $U$ and that for any projector, $P^2=P$. Note that the expression becomes independent of the unitary transformation, $U$, which also means that no transformation need take place between the two measurements (this will be important when we make a connection with temporal Bell's inequalities). The above probabilities, $p(n,m)$ allow us to define the following
\begin{equation}
\langle e^{\beta (W -\Delta F)} \rangle := \sum_{n,m} p(n,m) e^{\beta (E^i_n - E^f_m -\Delta F)} \; .
\end{equation}  
The average can now be evaluated to be
\begin{equation}
\langle e^{\beta (W -\Delta F)} \rangle = \sum_{n,m} tr  \{ P^f_m P^i_n\} \frac{e^{-\beta E^i_n}}{Z_i} e^{\beta (E^i_n - E^f_m)} \frac{Z_i}{Z_f} = \sum_m \frac{e^{-\beta E^f_m}}{Z_f} tr P^f_m = 1
\end{equation}  
which concludes the proof of the quantum version of the Jarzynski equality. 
\end{proof}

The real power of this equality is that it allows us to experimentally determine the free energy landscape of a system by repeatedly testing its ability to perform work (on average). This is because the quantity $\langle e^{\beta W}\rangle$ is experimentally accessible and, through Jarzynski, it is equal to $e^{\beta \Delta F}$ (which itself is a function of state only and therefore need not be averaged over). The free energy is, on the other hand, not easy to measure by any direct means in a typical complex system.  

Let us now specialize to a two level system, with eigen energies $\pm E$. The corresponding initial and final projectors will be labeled as $P^i_{\pm}$ and $P^f_{\pm}$ respectively.  It will again be convenient to use the Bloch representation for the projectors, $P^{i,f}_{\pm} = (1\pm s^{i,f}\cdot \sigma)/2$, where $s^{i,f}$ are the initial and the final measurement Bloch vectors and $s \cdot\sigma = s_x\sigma_x +s_y\sigma_y +s_z\sigma_z$ ($\sigma$s being the usual Pauli matrices). We then obtain
\begin{eqnarray}
\langle e^{\beta W} \rangle = \sum_{n,m} e^{\beta (E^f_m - E^i_n)} p(m,n) & = & \frac{e^{-\beta E}}{Z}tr P^f_+P^i_+ + \frac{e^{\beta E}}{Z} tr P^f_-P^i_- + \frac{e^{ \beta E}}{Z}tr P^f_+P^i_- + \frac{e^{-\beta E}}{Z}tr P^f_+P^i_+ \\
& = & \frac{1+s^i\cdot s^f}{2} (\frac{e^{ \beta E} + e^{-\beta E}}{Z}) + \frac{1-s^i\cdot s^f}{2} (\frac{e^{ \beta E} + e^{-\beta E}}{Z}) = 1 
\end{eqnarray} 
where $Z = 2\cosh \beta E$ is the (initial and final) partition function and $s^i\cdot s^f = s^i_x s^f_x +s^i_ys^f_y +s^i_zs^f_z$. This is an instructive calculation because it shows that even tough each of the four terms in the first line of the above depends on the measurement basis (i.e. on the angle between the two measurement directions, itself given by $s^i\cdot s^f$) the total average does not. This overall independence, of course, ``ensures" that the Jarzynski equality also holds identically in quantum mechanics (and not only classical physics). With this in mind, let us turn to analysing different work moments. It is clear that spatial entanglement, which is already implicitly encoded into above states, makes no extra contribution to the discussion. The impact of temporal entanglement, on the other hand, is more subtle.

\section{Expansion of work in terms of temporal entanglement}

Having defined the quantum protocol for thermodynamical average work, and related it to free energy, we now proceed to look at various moments on the exponential of work. By using the Taylor expansion we have 
\begin{equation}
\langle e^{\beta W}\rangle = \sum_{k=0}^{\infty} \frac{\beta^{k}}{k!} \langle W^{k} \rangle \; ,
\end{equation}
which allows us to look at the individual moments of work, $W$. Let us for simplicity again consider a protocol involving a qubit, whose initial and the final energies are the same ($\pm E$). This also makes the corresponding free energies equal to each other. We start in a thermal state in the eigen-basis of an operator $A$ and then, following a trivial unitary transformation $U=I$, we will measure in the eigen-basis of $B$ (which, as we said, will be the same as the eigen-basis of $U$).   

\begin{itemize}
\item The first term is the average work is given by 
\begin{equation}
\langle W_{A,B} \rangle = \sum_{n,m} (E_n - E_m) p(n,m) = -E \tanh (\beta E)(1 - s^i \cdot s^f) \label{work} \; .
\end{equation}  
In order to show that there is a difference between the quantum and classical thermodynamical work, 
we consider a scenario where we make two initial measurements, $A_1$ and $A_2$ and two final measurements $B_1$ and $B_2$. We then construct a (Bell-CHSH-like) combination of the corresponding amounts of work: 
\begin{equation} 
\langle W \rangle_{{\cal B}} := \langle W_{A_1,B_1}\rangle + \langle W_{A_1,B_2}\rangle + \langle W_{A_2,B_1}\rangle - \langle W_{A_2,B_2}\rangle \; .
\end{equation}
Substituting eq. (\ref{work}) into above we obtain
\begin{equation}
\langle W \rangle_{{\cal B}} = -E \tanh (\beta E)(2 - (s_1^i\cdot s_1^f + s_1^i\cdot s_2^f + s_2^i\cdot s_1^f - s_2^i\cdot s_2^f)) \; ,
\end{equation}
where $s_{1,2}^i$ are the Bloch vectors defining the eigen-basis of $A_{1,2}$ and $s_{1,2}^f$ are the Bloch vectors defining the eigen-basis of $B_{1,2}$. 
It is clear that the last expression in brackets is the same as in the temporal Bell inequality (cf. eq. (\ref{quantum-Bell})). As soon as the absolute value of that quantity exceeds the value of $2$, we have a genuinely quantum behaviour. For instance, the classical average work cannot be positive in this case, though quantumly the value can be as high as $2E\tanh (\beta E) (\sqrt{2}-1)$. Note that this quantity vanishes with the increasing temperature, which is intuitively pleasing since the high temperature limit is usually thought of as classical. This particular feature is in contrast to the temporal Bell inequalities, which themselves are independent of the initial state. 

\item We now investigate the second order term. The quantity of interest is the average work squared which is, by definition, equal to  
\begin{equation}
\langle W^2_{A,B}\rangle = \sum_{nm} p_{A,B}(n,m) (E_n - E_m)^2 = 2E^2 (1-s^i \cdot s^f) 
\end{equation}
We then make two initial and two final measurements to construct the following (again CHSH-like) quantity
\begin{equation} 
\langle W^2 \rangle_{{\cal B}} := \langle W^2_{A_1,B_1}\rangle + \langle W^2_{A_1,B_2}\rangle + \langle W^2_{A_2,B_1}\rangle - \langle W^2_{A_2,B_2}\rangle \; .
\end{equation}
Analogous expressions are obtained for the other three terms and putting it all together we have 
\begin{equation}
\langle W^2 \rangle_{{\cal B}} = 2E^2 (2 - (s_1^i \cdot s_1^f + s_1^i \cdot s_2^f + s_2^i \cdot s_1^f - s_2^i \cdot s_2^f)) \; .
\end{equation}
It is easily recognised that the last part of this expressions is again the temporal Bell inequality. Therefore, classically
\begin{equation}
0 \leq \langle W^2 \rangle_{{\cal B},classical} \leq 8E^2 
\end{equation}
Quantumly, on the other hand, we have seen that this inequality can be violated and we can have 
\begin{equation}
\langle W^2 \rangle_{{\cal B},quantum} = 4(1+\sqrt{2})E^2 
\end{equation}
Note that, unlike in the first moment, here the difference between quantum and classical behaviour does not depend on temperature. This is because the second moment of work is independent of the initial state of the system much like the temporal Bell inequality. It soon becomes clear that all the even moments have exactly the same feature of the initial state independence, while all the odd ones behave like the first moment. We can therefore naturally write the general expression for the $n$-th order.     

\item The general expression for the $n$-th moment of work is (as is easily seen from the first two moments)
\begin{equation}
\langle W^n \rangle  = 2^{n-1} E^n (1-s^i \cdot s^f) \frac{e^{-\beta E} + (-1)^n e^{+\beta E}}{e^{-\beta E} + e^{+\beta E}} \; .
\end{equation}
Therefore, all orders can lead to similar CHSH-like temporal inequalities, within the appropriate setting involving two initial and two final measurements. 

\end{itemize}

Following the above logic, we conclude that all of the work moments do, in principle, display different classical in quantum behaviour. Remarkably, however, as can be seen from the Jarzynski inequality, the sum of all orders cancels out all quantum features and becomes measurement independent (and equal to the free energy difference between the initial and the final states, which is, in our case, identically equal to zero). This is all the more remarkable since, as we have seen, all even moments of work are, in fact, independent of the initial state and therefore the difference between quantum and classical behaviour persists at any temperature. We now proceed to show formally that the summation of all orders simply becomes a unity (as the Free energy difference is zero in this case):
\begin{eqnarray}
\langle e^{\beta W}\rangle = \sum_{k} \frac{\beta^{k}}{k!} \langle W^{k}\rangle & = & \frac{(1-s^i \cdot s^f)}{2}\times \left\{- \sum_{k\in even} \frac{(2\beta E)^{k}}{k!} \tanh (\beta E) + \sum_{k\in odd} \frac{(2\beta E)^{k}}{k!} \right\}\\
& = & \frac{(1-s^i \cdot s^f)}{2} (\cosh^2 (2\beta E) - \sinh^2 (2\beta E) -1) + 1 = 1  
\end{eqnarray}
where we have used the Taylor expansion for the hyperbolic sinus and cosines functions. Thus, even though each term in the expansion demonstrates clearly a difference between the quantum and classical bahaviour, the total sum does not. Averaging this over two initial and two final settings does not change anything. We now proceed to discuss four important points.

\section{Discussion and conclusions}

The first important point to note is that the thermodynamical setting we presented above is slightly different to the temporal Bell's inequalities setting. This is because the initial state in our work scenario depends on the first observable to be measured ($A_1$ or $A_2$), while in temporal Bell's inequalities, the initial state is always the same (though changing it would not affect the result which is as we stressed independent of it). Our formulation of the work - temporal entanglement relationship is, therefore, somewhat more akin to the non-locality analysis in the work of Williamson et al \cite{Mark}. There, the authors showed how to achieve spatial Bell violation but with (spatially) separable initial states of two qubits (where each state was made dependent on the particular subsequent measurements to be executed). However, this turns out not to be a fundamental problem for us. In order to see why, we can instead use the original Bell inequality with only three (instead of four) settings. Then, the first measurement is always $A$, while the latter two can either be $B_1$ or $B_2$. The Bell inequality is then given by $1+\langle B_1B_2\rangle\geq |\langle AB_1\rangle  - \langle AB_2\rangle|$. In this case, the initial state is always fixed to be the thermal state in the basis of $A$. We then always measure first the observable $A$ and then make a random choice between $B_1$ and $B_2$. The violation of this Bell inequality leads immediately to a higher amount of quantum thermodynamical work. 

Secondly, we stressed the fact that even though all individual moments differ quantumly from their corresponding classical value, the sum does not. This is somewhat reminiscent of Feynman's idea to use negative probabilities to ``explain" the quantum violation of Bell's inequalities \cite{Feynman}. Briefly, his logic is that we need not give up either locality or reality assumptions (or the implicitly assumed freedom of will) that are behind deriving Bell's inequalities, as long as we are prepared to introduce the possibility of having negative probabilities. Feynman argued that negative probabilities can be justified by their calculational utility in the same way as negative numbers (minus three sheep makes little sense in real life, but its use is still a convenient way of expressing debt). Similarly to the Jarzynski equality, in Feynman's treatment, the negative probabilities are only used mid-point in derivations, and the final result never has any negative probabilities. Here in our case ``classical" behaviour is recovered from an infinite series of ``quantum" terms. 

Thirdly, we would like to discuss the present work in relation to thermodynamical witnesses of (spatial) quantum entanglement, especially quantities such as the internal energy, heat capacity and magnetic susceptibility \cite{Amico}. First, we need a clarification. How can thermodynamical quantities witness anything like entanglement? All thermodynamic quantities are derivatives of the free energy, which itself is dependent only on the partition function. But, the partition function in turn depends only on the energy eigenvalues (and not eigenstates), so how can it contain any information regarding entanglement? We have already discussed this issue elsewhere \cite{Amico} and the key is that the quantum partition function still contains the information about the eigen-basis through the exponential of the Hamiltonian, $e^{-\beta H}$. The second derivative of this quantity, say with respect with $\beta$, will give us a two-point correlation function (and possibly higher order correlations, depending on the Hamiltonian), effectively resulting in the expression for heat capacity. Our temporal correlations reflect quantum correlations for exactly the same reason. What they show, however, in contrast to spatial correlations, is that the first and the second moments fix all other moments. This is due to the property of measurements in quantum physics which erase all dependence on the past (in other words, projective measurements are Markovian). Namely, once we project a given system onto a particular state, that state then contains the only relevant information needed to describe for all future behaviour (it makes no difference what the state was that existed before the measurement was performed). Therefore, there are no genuine three point (or higher) quantum correlations in time - they can all be reduced to two point correlations.      

Our fourth comment is on the thermodynamical arrow of time. It is perhaps the most baffling, at least at first sight. The logic of Jarzynski is sometimes used to talk about the thermodynamical arrow of time, in the sense that there is a difference between probabilities in the forward and backward directions. To see this, imagine that we start the system in the state $\rho_f$, then perform a measurement in the $P^f_m$ basis, followed by the evolution $U^{-1} = e^{iH_i t}$. Finally a measurement is performed in the basis $P^i_n$. The probability $p(n,m)$ in this backward direction is now different from the one in the forward direction (their ratio is given by the Crooks relation \cite{Crooks}). As far as Bell's correlations are concerned, there is no such distinction between the past and future. It makes no difference which measurement is made first and which second, the resulting correlations are identical. How can this be? Namely, how can the Jarzynski equality reveal a time-asymmetry while the temporal Bell inequality is blind to it? The answer is that in the Jarzynski scenario, going in the reverse direction implies starting from a different thermal state (i.e. the state $\rho_f=\exp\{{-\beta H_f}\}/Z_f$, which is diagonal in the eigen-basis of $H_f$). Thus the probabilities to go in the forward and backward directions differ exactly by the ratio of the probabilities to be in the eigen-states of
$\rho_i$ and $\rho_f$ respectively. This is why, if we assume the same spectra for the initial and the final state, this distinction vanishes and the thermodynamical work and Bell inequalities can be directly related. The overarching logic is that the thermodynamical arrow of time is simply due to a special initial condition (see e.g. \cite{Price}), and does not stem from any dynamical time asymmetry (no such asymmetry exist in quantum physics even in measurement correlations). A simple way of understanding this is to imagine that measurements are represented as unitary couplings between the system and an apparatus. Any such unitary coupling could, by definition of unitarity, be reversed. However, this reversal implies that when going backwards we need to start from the same state of the system and the apparatus that was the end result of the unitary coupling executing the measurement (the same conclusion is reached by using the formalism in \cite{Aharonov}). If, for some reason, we change the state of the system (like we do when calculating the backward probability in Jarzynski) we can no longer reverse the effect of the measurement. This is all there is to thermodynamical irreversibility, and it is why temporal entanglement reflects no such asymmetry.    

Finally, we mention that our connection between temporal entanglement and thermodynamical work can easily be tested experimentally by using ideas already proposed in \cite{Huber} whose aim was to test the quantum version of the Jarzynski equality. The authors suggest utilising the vibrational degrees of freedom of a trapped ion as the quantum system and making measurements of phonons at two different times, separated by a coherent unitary evolution in between. It is likely that in the near future such or some related set up could also be used for purposes of testing the ideas presented in the present paper. Alternatively, the present paper shows that whoever decides to experimentally test temporal entanglement is also (though perhaps unwittingly) testing a special form of the quantum Jarzynski equality since the two are effectively the same.    

\textit{Acknowledgments}: The author acknowledges financial support from the Templeton foundation, Leverhulme Trust (UK), as well as the National Research Foundation and Ministry of Education, in Singapore. The author is a fellow of Wolfson College Oxford.

\end{document}